\documentclass[11pt,a4paper]{article}
\usepackage{amsfonts}
\usepackage{amsmath,amssymb}
\usepackage{epsfig,graphicx}

\input{tcilatex}

\begin{document}

\begin{center}
{\huge Emergent SUSY Theories: QED, SM \& GUT}

{\Huge \bigskip \bigskip \bigskip }

\textbf{J.L.~Chkareuli}$^{1,2}$

$^{1}$\textit{Center for Elementary Particle Physics, Ilia State University,
0162 Tbilisi, Georgia\ \vspace{0pt}\\[0pt]
}

$^{2}$\textit{E. Andronikashvili} \textit{Institute of Physics, 0177
Tbilisi, Georgia\ }

\textit{\bigskip }

\bigskip

\bigskip

\bigskip

\bigskip

\textbf{Abstract}

\bigskip

\bigskip
\end{center}

It might be expected that only global symmetries are fundamental symmetries
of Nature, whereas local symmetries and associated massless gauge fields
could solely emerge due to spontaneous breaking of underlying spacetime
symmetries involved, such as relativistic invariance and supersymmetry. This
breaking, taken in the form of the nonlinear $\sigma $-model type pattern
for vector fields or superfields, puts essential restrictions on geometrical
degrees of freedom of a physical field system that makes it to adjust itself
in such a way that its global internal symmetry $G$ turns into the local
symmetry $G_{loc}$. Remarkably, this emergence process may naturally be
triggered by spontaneously broken supersymmetry, as is illustrated in detail
by an example of a general supersymmetric QED model which is then extended
to electroweak models and grand unified theories. Among others, the $%
U(1)\times SU(2)$ symmetrical Standard Model and flipped $SU(5)$ GUT appear
preferable to emerge at high energies.

\bigskip

\bigskip

\bigskip

\bigskip

\bigskip \bigskip

\bigskip

\bigskip

\bigskip\bigskip

\bigskip

\bigskip

\bigskip

\begin{center}
{\tiny Invited talk at the International Workshop \textquotedblright What
Comes Beyond the Standard Model?\textquotedblright\ (21-28 July 2014, Bled,
Slovenia)}
\end{center}

{\tiny \bigskip }

%%%%%%%%%%%%%%%%%%%%%%%%%%%%%%%%%%%%%%%%%%%%%%%%%%%%%%%
\thispagestyle{empty}\newpage

\section{Introduction}

We all believe that internal gauge symmetries form the basis of modern
particle physics being most successfully realized within the celebrated
Standard Model (SM) of quarks and leptons and their fundamental strong, weak
and electromagnetic interactions. At the same time, local gauge invariance,
contrary to a global symmetry case, may look like a cumbersome geometrical
input rather than a "true" physical principle, especially in the framework
of an effective quantum field theory\ (QFT) becoming, presumably, irrelevant
at very high energies. In this connection, one could wonder whether there is
any basic dynamical reason that necessitates gauge invariance and the
associated masslessness of gauge fields as some emergent phenomenon arising
from a more profound level of dynamics. By analogy with a dynamical origin
of massless scalar particle excitations, which is very well understood in
terms of spontaneously broken global internal symmetries \cite{NJL}, one
could think that the origin of massless gauge fields as vector
Nambu-Goldstone (NG) bosons are related to the spontaneous Lorentz
invariance violation (SLIV) that is the minimal spacetime global symmetry
underlying particle physics. This well-known approach providing a viable
alternative to quantum electrodynamics \cite{bjorken}, gravity \cite{ph} and
Yang-Mills theories \cite{eg} has a long history started over fifty years
ago, though has been significantly revised in the recent years \cite{cfn,
jb, kraus, bluhm}.

\subsection{An emergence conjecture}

Directly or indirectly, the approach mentioned includes several key points
which in a conventional QFT framework may be formulated nowadays in the
following way (see \cite{jlc} and comprehensive references therein):

\begin{itemize}
\item Only global symmetries are fundamental symmetries of Nature. Local
symmetries and associated massless gauge vector (tensor) fields could only
emerge due to some phase transition producing them as appropriate
Nambu-Goldstone modes,

\item The underlying Lorentz invariance is proposed to be spontaneously
broken since only spacetime symmetry breaking could basically provide an
existence of vector (tensor) emerging modes which mediate all interactions
involved,

\item The theory itself is proposed to be "physically" viable in the sense
that any appropriate initial value condition (IVC), which determines the
subsequent dynamical evolution of a physical field system, is uniquely
satisfied. This means in turn that an interacting field system can not be
superfluously restricted in the number of physical degrees of freedom in
order to remain physical,

\item Together, they naturally lead to the \textbf{\ gauge symmetry
emergence }(GSE)\textbf{\ }conjecture which I will follow throughout the
paper: \textit{Let there be given an interacting field system containing
some vector field (or vector field multiplet) }$A_{\mu }$\textit{\ together
with fermion (}$\psi $), \textit{scalar (}$\phi $\textit{) and other matter
fields in an arbitrary relativistically invariant Lagrangian\ }$L(A_{\mu
},\psi ,\phi ,...)$\textit{\ which possesses only global Abelian or
non-Abelian internal symmetry }$G$\textit{. Suppose that an underlying
relativistic invariance of this field system is spontaneously broken in
terms of t\textit{he} "length-fixing" covariant constraint put on vector
fields,}%
\begin{equation}
A_{\mu }A^{\mu }=n^{2}\mathrm{M}^{2}  \label{const}
\end{equation}%
\textit{(where }$\mathrm{M}$\textit{\ stands for the proposed SLIV scale,
while }$n_{\mu }$\textit{\ is a properly-oriented unit Lorentz vector, }$%
n^{2}=n_{\mu }n^{\mu }=\pm 1$\textit{)}. \textit{If this constraint is
preserved under the time development given by the field equations of motion,
then in order to be\textit{\ protected }from further reduction in degrees of
freedom this system will modify its global symmetry }$G$\textit{\ into a
local symmetry }$G_{loc},$ \textit{that will in turn convert the vector
field constraint itself into a gauge condition thus virtually resulting in
gauge invariant and Lorentz invariant theory.}
\end{itemize}

To see how technically a global internal symmetry may be converted into a
local one, let us consider in some detail the question of consistency of a
possible constraint for a general 4-vector field $A_{\mu }$ with its
equation of motion in an Abelian symmetry case, $G=U(1)$. In the presence of
the SLIV constraint $C(A)=A_{\mu }A^{\mu }-n^{2}\mathrm{M}^{2}=0$ (\ref%
{const}), it follows that the equations of motion can no longer be
independent. The important point is that, in general, the time development
would not preserve the constraint. So the parameters in the starting
Lagrangian have to be adjusted in such a way that effectively we have one
less equation of motion for the vector field $A_{\mu }$ not to be
superfluously restricted. This means that there should be some relationship
given by a functional equation $F(C=0;$ \ $E_{A},E_{\psi },...)=0$ between
all the vector and matter field Eulerians involved\footnote{%
The field Eulerians ($E_{A}$, $E_{\psi }$, ...) are determined, as usual, $%
(E_{A})^{\mu }\equiv \partial L/\partial A_{\mu }-\partial _{\nu }[\partial
L/\partial (\partial _{\nu }A_{\mu })]$, and so forth.} which are
individually satisfied on the mass shell. According to Noether's second
theorem \cite{net} such a relationship gives rise to an emergence of local
symmetry for the field system considered provided that the functional $F$
satisfies the same symmetry requirements of Lorentz and translational
invariance, as well as all the global internal symmetry requirements, as the
general starting Lagrangian does.

In this way, the nonlinear SLIV condition (\ref{const}), due to which true
vacuum in the theory is chosen and massless gauge fields are generated, may
provide a dynamical setting for all underlying internal symmetries involved
through the GSE conjecture \cite{jlc}. One might think that the
length-fixing vector field constraint (\ref{const}) itself first introduced
by Nambu in a conventional QED framework \cite{nambu} (for some extensions
and generalizations, see also \cite{az, kep, jej, urr, gra, cfn2}) does not
especially stand out in the present context. Actually, it seems that the GSE
conjecture might be equally formulated for any type of covariant constraint,
say for the spin-1 vector field condition, $\partial _{\mu }A^{\mu }=0$ \cite%
{ogi3}. However, as is generally argued in \cite{jlc}, the SLIV constraint (%
\ref{const}) appears to be the only one whose application leads to a full
conversion of an internal global symmetry $G$\ into a local symmetry $%
G_{loc} $ that forces a given field system to remain always physical. Other
constraints could only lead to partial gauge invariance being broken by some
terms in an emerging theory.

Based upon the SLIV constraint (\ref{const}), the starting vector field $%
A_{\mu }$ may be expanded around the true vacuum configuration in the
theory, 
\begin{equation}
A_{\mu }=a_{\mu }+n_{\mu }\sqrt{\mathrm{M}^{2}-n^{2}a^{2}}\text{ , \ }n_{\mu
}a_{\mu }=0\text{ \ }(a^{2}\equiv a_{\mu }a^{\mu })\text{ ,}  \label{vev1}
\end{equation}%
which means that it develops the vacuum expectation value (VEV) $\langle
A_{\mu }\rangle =n_{\mu }\mathrm{M}$. Meanwhile, its $a_{\mu }$ components
which are orthogonal to the Lorentz violating direction $n_{\mu }$ describe
a massless vector NG boson being an eventual gauge field (photon) candidate.

\subsection{Gauge invariance \textit{versus} spontaneous Lorentz violation}

One can see that the gauge theory framework, be it taken from the outset or
emerged, makes in turn spontaneous Lorentz violation to be physically
unobservable both in Abelian and non-Abelian symmetry case. In substance,
the essential part of the SLIV pattern (\ref{vev1}), due to which the vector
field $A_{\mu }(x)$ develops the VEV $\mathrm{M}$, may itself be treated as
a pure gauge transformation with a gauge function linear in coordinates, $%
\omega (x)=$ $n_{\mu }x^{\mu }\mathrm{M}$. This is what one could refer to
as the generic non-observability of SLIV in gauge invariant theories. I
shall call it the "inactive" SLIV in contrast to the "active" SLIV case
where physical Lorentz invariance could effectively occur. From the present
standpoint, the only way for an active SLIV to occur would be if emergent
gauge symmetries presented above were slightly broken at small distances.
This could inevitably happen, for example, in a partially gauge invariant
theory which might appear if the considered field system could become "a
little unphysical" at distances being presumably controlled by quantum
gravity \cite{par}. One may think that quantum gravity could in principle
hinder the setting of the required IVC in the appropriate Cauchy problem
(thus admitting a superfluous restriction of vector fields) due to the
occurrence of some gauge-noninvariant high-order operators near the Planck
scale. As a consequence, through special dispersion relations appearing for
matter and gauge fields, one is led a new class of phenomena which could be
of distinctive observational interest in particle physics and astrophysics.
They include a significant change in the Greizen-Zatsepin-Kouzmin cutoff for
ultra-high energy cosmic-ray nucleons, stability of high-energy pions and $W$
bosons, modification of nucleon beta decays, and some others just in the
presently accessible energy area in cosmic ray physics \cite{par} (for many
phenomenological aspects, see pioneering works \cite{2, 3}).

\subsection{SUSY profile of emergent theories}

The role of Lorentz invariance may change, and its spontaneous violation may
not be the only reason why massless photons and other gauge fields could
dynamically appear, if spacetime symmetry is further enlarged. In this
connection, special interest is related to supersymmetry which has made a
serious impact on particle physics in the last decades (though has not been
yet discovered). Actually, as we will see, the situation is changed
dramatically in the SUSY inspired emergent gauge theories. In sharp contrast
to non-SUSY analogs, it appears that the spontaneous Lorentz violation
caused by an arbitrary potential of vector superfield $V(x,\theta ,\overline{%
\theta })$ never goes any further than some nonlinear gauge condition put on
its vector field component $A_{\mu }(x)$ associated with a photon or any
other gauge field. Remarkably, this condition coincides, as we shall see
below, with\ the SLIV constraint (\ref{const}) given above in the GSE
conjecture. This allows to think that physical Lorentz invariance is
somewhat protected by SUSY, thus only requiring the "condensation" of the
gauge degree of freedom in the vector field $A_{\mu }$. The point is,
however, that even in the case when SLIV is not physical it inevitably leads
to the generation of massless photons as vector NG bosons provided that SUSY
itself is spontaneously broken. In this sense, a generic trigger for
massless photons to dynamically emerge happens to be spontaneously broken
supersymmetry rather than physically manifested Lorentz noninvariance.

While there are many papers in the literature on Lorentz noninvariant
extensions of supersymmetric models (for some interesting ideas, see \cite%
{bs, pos1} and references therein), an emergent gauge theory in a SUSY
context has only recently been introduced \cite{jlc, c}. Actually, the
situation was shown to be seriously changed in a SUSY context which
certainly disfavors some emergent models considered above. It appears that,
while the constraint-based models of an inactive SLIV successfully matches
supersymmetry, the composite and potential-based models of an active SLIV
leading to physical Lorentz violation cannot be conceptually realized in the
SUSY context. The reason is that, in contrast to an ordinary vector field
theory where all kinds of polynomial terms $(A_{\mu }A^{\mu })^{n}$ ($%
n=1,2,...$) can be included into the Lagrangian in a Lorentz invariant way,
SUSY theories only admit the bilinear mass term $A_{\mu }A^{\mu }$ in the
vector field potential energy. As a result, without a stabilizing
high-linear (at least quartic) vector field terms, the potential-based SLIV
never occurs in SUSY theories. The same could be said about composite models 
\cite{bjorken, ph, eg} as well: a fundamental Lagrangian with multi-fermi
current-current interactions can not be constructed from any matter chiral
superfields. So, all the models mentioned above, but the constraint-based
models determined by the GSE conjecture (\ref{const}), are ruled out in the
SUSY framework and, therefore, between the two basic SLIV versions, active
and inactive, SUSY unambiguously chooses the inactive SLIV case.

\subsection{Outline of the paper}

The paper is organized in the following way. In the next section 2 I
consider supersymmetric QED model extended by an arbitrary polynomial
potential of massive vector superfield that breaks gauge invariance in the
SUSY invariant phase. However, the requirement of vacuum stability in such
class of models makes both supersymmetry and Lorentz invariance to become
spontaneously broken. As a consequence, the massless photino and photon
appear as the corresponding Nambu-Goldstone zero modes in an emergent SUSY
QED, and also a special gauge invariance is simultaneously generated. Due to
this invariance all observable relativistically noninvariant effects appear
to be completely cancelled out and physical Lorentz invariance is recovered.
Further in section 3, all basic arguments developed in SUSY QED are
generalized successively to the Standard Model and Grand Unified Theories
(GUTs). For definiteness, I focus on the $U(1)\times SU(N)$ symmetrical
theories. Such a split group form is dictated by the fact that in the pure
non-Abelian symmetry case one only has the SUSY invariant phase in the
theory that makes it inappropriate for an outgrowth of an emergence process.
As possible realistic realizations, the Standard Model case with the \
electroweak $U(1)\times SU(2)$ symmetry and flipped $SU(5)$ GUT including
some immediate applications are briefly discussed. And finally in section 4,
I summarize the main results and conclude.

The present talk is complimentary to my last year talk in Bled \cite{lec}.
Some more detail can also be found in the recent extended paper \cite{jlc}.

\section{Emergent SUSY theories: a QED primer}

In contrast to attempts simply probing physical Lorentz noninvariance
through some SM extensions \cite{bluhm, 2} with hypothetical external vector
(tensor) field backgrounds originated around the Planck scale, we will
principally focus here on a spontaneous Lorentz violation in an ordinary
Standard Model framework itself. Particularly, we will try to extend an
emergent SM with electroweak bosons appearing as massless vector NG modes to
their supersymmetric analogs \cite{jlc, c}. Such theories seem to open a new
avenue for exploring the origin of gauge symmetries. Indeed, as I discussed
at the previous workshop \cite{lec}, the emergent SUSY theories, in contrast
to the non-SUSY ones, could naturally have some clear observational
signature. Actually, we have seen above that ordinary emergent gauge
theories are physically indistinguishable from the conventional ones unless
gauge invariance becomes broken being caused by some high-dimension
couplings. Meanwhile, their SUSY counterparts - supersymmetric QED, SM and
GUT - can be experimentally verified in another way. The point is that they
generically emerge only if supersymmetry is spontaneously broken in a
visible sector in order to ensure stability of the underlying theory.
Therefore, the verification of emergent theories is now related to an
inevitable emergence of a goldstino-like photino state in the SUSY particle
spectrum at low energies, while physical Lorentz invariance may be still
left intact.

\subsection{Spontaneous SUSY violation}

Since gauge invariance is not generically assumed in an emergent approach,
all possible gauge-noninvariant couplings could in principle occur in the
theory in a pre-emergent phase. The most essential couplings, as I discussed
earlier \cite{lec}, appear to be the vector field self-interaction terms
triggering an emergence process in non-SUSY theories. Starting from this
standpoint, I consider a conventional supersymmetric QED being similarly
extended by an arbitrary polynomial potential of \ a general vector
superfield $V(x,\theta ,\overline{\theta })$ which in the standard
parametrization \cite{wess} has a form 
\begin{eqnarray}
V(x,\theta ,\overline{\theta }) &=&C+i\theta \chi -i\overline{\theta }%
\overline{\chi }+\frac{i}{2}\theta \theta S-\frac{i}{2}\overline{\theta }%
\overline{\theta }S^{\ast }  \notag \\
&&-\theta \sigma ^{\mu }\overline{\theta }A_{\mu }+i\theta \theta \overline{%
\theta }\overline{\lambda ^{\prime }}-i\overline{\theta }\overline{\theta }%
\theta \lambda ^{\prime }+\frac{1}{2}\theta \theta \overline{\theta }%
\overline{\theta }D^{\prime },  \label{par1}
\end{eqnarray}%
where its vector field component $A_{\mu }$ is usually associated with a
photon. Note that, apart from an ordinary photino field $\lambda $ and an
auxiliary $D$ field, the superfield (\ref{par1}) contains in general some
additional degrees of freedom in terms of the dynamical $C$ and $\chi $
fields and nondynamical complex scalar field $S$ (I have used the brief
notations, $\lambda ^{\prime }=\lambda +\frac{i}{2}\sigma ^{\mu }\partial
_{\mu }\overline{\chi }$ \ and $D^{\prime }=D+\frac{1}{2}\partial ^{2}C$
with $\sigma ^{\mu }=(1,\overrightarrow{\sigma })$ and $\overline{\sigma }%
^{\mu }=(1,-\overrightarrow{\sigma })$). The corresponding Lagrangian can be
written as%
\begin{equation}
L=L_{SQED}+\frac{1}{2}D^{2}+\sum_{k=1}b_{k}V^{k}|_{D}  \label{slag}
\end{equation}%
where, besides a standard SUSY QED part, new potential terms are presented
in the sum by corresponding $D$-term expansions $V^{k}|_{D}$ of the vector
superfield (\ref{par1}) into the component fields ($b_{k}$ are some
constants). It can readily be checked that the first term in this expansion
is the known Fayet-Iliopoulos $D$-term, while other terms only contain
bilinear, trilinear and quartic combination of the superfield components $%
A_{\mu }$, $S$, $\lambda $ and $\chi $, respectively.

Actually, the higher-degree terms only appear for the scalar field component 
$C(x)$. Expressing them all in terms of the $C$ field polynomial%
\begin{equation}
P(C)=\sum_{k=1}\frac{k}{2}b_{k}C^{k-1}(x)  \label{pot}
\end{equation}%
and its first three derivatives 
\begin{equation}
P_{C}^{\prime }\equiv \frac{\partial P}{\partial C}\text{ , \ \ }%
P_{C}^{\prime \prime }\equiv \frac{\partial ^{2}P}{\partial C^{2}}\text{ , \
\ }P_{C}^{\prime \prime \prime }\equiv \frac{\partial ^{3}P}{\partial C^{3}}%
\text{ }  \label{dd}
\end{equation}%
one has for the whole Lagrangian $L$ 
\begin{eqnarray}
L &=&L_{SQED}+\frac{1}{2}D^{2}+\text{ }P\left( D+\frac{1}{2}\partial
^{2}C\right)  \notag \\
&&+P_{C}^{\prime }\left( \frac{1}{2}SS^{\ast }-\chi \lambda ^{\prime }-%
\overline{\chi }\overline{\lambda ^{\prime }}-\frac{1}{2}A_{\mu }A^{\mu
}\right)  \notag \\
&&+\text{ }\frac{1}{2}P_{C}^{\prime \prime }\left( \frac{i}{2}\overline{\chi 
}\overline{\chi }S-\frac{i}{2}\chi \chi S^{\ast }-\chi \sigma ^{\mu }%
\overline{\chi }A_{\mu }\right) +\frac{1}{8}P_{C}^{\prime \prime \prime
}(\chi \chi \overline{\chi }\overline{\chi })\text{ .}  \label{lag3}
\end{eqnarray}%
As one can see, extra degrees of freedom related to the $C$ and $\chi $
component fields in a general vector superfield $V(x,\theta ,\overline{%
\theta })$ appear through the potential terms in (\ref{lag3}) rather than
from the properly constructed supersymmetric field strengths, as appear for
the vector field $A_{\mu }$ and its gaugino companion $\lambda $.

Note that all terms in the sum in (\ref{slag}) except Fayet-Iliopoulos $D$%
-term\ explicitly break gauge invariance. However, as we will see later in
this section, the special gauge invariance constrained by some gauge
condition will be recovered in the Lagrangian in the broken SUSY phase.
Furthermore, as is seen from (\ref{lag3}), the vector field $A_{\mu }$ may
only appear with bilinear mass term in the polynomially extended superfield
Lagrangian (\ref{slag}) in sharp contrast to the non-SUSY theory case where,
apart from the vector field mass term, some high-linear stabilizing\ terms
necessarily appear in a similar polynomially extended Lagrangian. This means
in turn that physical Lorentz invariance is still preserved. Actually, only
supersymmetry appears to be spontaneously broken in the theory.

Indeed, varying the Lagrangian $L$ with respect to the $D$ field we come to 
\begin{equation}
D=-P(C)  \label{d}
\end{equation}%
that finally gives the following potential energy for the field system
considered 
\begin{equation}
U(C)=\frac{1}{2}[P(C)]^{2}\text{ .}  \label{pot1}
\end{equation}%
The potential (\ref{pot1}) may lead to spontaneous SUSY breaking in the
visible sector provided that the polynomial $P$ (\ref{pot}) has no real
roots, while its first derivative has, 
\begin{equation}
P\neq 0\text{ , \ }P_{C}^{\prime }=0.\text{\ }  \label{der}
\end{equation}%
This requires $P(C)$ to be an even degree polynomial with properly chosen
coefficients $b_{k}$ in (\ref{pot}) that will force its derivative $%
P_{C}^{\prime }$ to have at least one root, $C=C_{0}$, in which the
potential (\ref{pot1}) is minimized. Therefore, supersymmetry is
spontaneously broken and the $C$ field acquires the VEV 
\begin{equation}
\left\langle C\right\rangle =C_{0}\text{ , \ }P_{C}^{\prime }(C_{0})=0\text{
.}  \label{vvv}
\end{equation}%
As an immediate consequence, that one can readily see from the Lagrangian $L$
(\ref{lag3}), a massless photino $\lambda $ being Goldstone fermion in the
broken SUSY phase make all the other component fields in the superfield $%
V(x,\theta ,\overline{\theta })$ including the photon to also become
massless. However, the question then arises whether this masslessness of the
photon will be stable against radiative corrections since gauge invariance
is explicitly broken in the Lagrangian (\ref{lag3}). I show below that it
could be the case if the vector superfield $V(x,\theta ,\overline{\theta })$
would appear properly constrained.

\subsection{Instability of superfield polynomial potential}

Let us first analyze possible vacuum configurations for the superfield
components in the polynomially extended QED case taken above. In general,
besides the "standard" potential energy expression (\ref{pot1}) determined
solely by the scalar field component $C(x)$ of the vector superfield (\ref%
{par1}), one also has to consider other field component contributions into
the potential energy. A possible extension of the potential energy (\ref%
{pot1}) seems to appear only due to the pure bosonic field contributions,
namely due to couplings of the vector and auxiliary scalar fields, $A_{\mu }$
and $S$, in (\ref{lag3}) 
\begin{equation}
U_{tot}=\frac{1}{2}P^{2}+\frac{1}{2}P_{C}^{\prime }\left( A_{\mu }A^{\mu
}-SS^{\ast }\right) \text{ }  \label{pot1a}
\end{equation}%
rather than due to the potential terms containing the superfield fermionic
components. It can be immediately seen that these new couplings in (\ref%
{pot1a}) can make the potential unstable since the vector and scalar fields
mentioned may in general develop any arbitrary VEVs. This happens, as
emphasized above, due the fact that their bilinear term contributions are
not properly compensated by appropriate four-linear field terms which are
generically absent in a SUSY theory context.

\subsection{Stabilization of vacuum by constraining vector superfield}

The only possible way to stabilize the theory seems to seek the proper
constraints on the superfield component fields ($C$, $A_{\mu }$, $S$)
themselves rather than on their expectation values. This will be done again
through some invariant Lagrange multiplier coupling simply adding its $D$
term to the above Lagrangian (\ref{slag}, \ref{lag3}) 
\begin{equation}
L_{tot}=L+\frac{1}{2}\Lambda (V-C_{0})^{2}|_{D}\text{ ,}  \label{ext}
\end{equation}%
where $\Lambda (x,\theta ,\overline{\theta })$ is some auxiliary vector
superfield, while $C_{0}$ is the constant background value of the $C$ field
which minimizes the potential $U$ (\ref{pot1}). Accordingly, the potential
vanishes for the supersymmetric minimum or acquires some positive value
corresponding to the SUSY breaking minimum (\ref{der}) in the visible
sector. I shall consider both cases simultaneously using the same notation $%
C_{0}$ for either of the background values of the $C$ field.

Writing down the Lagrange multiplier $D$ term in (\ref{ext}) through the
component fields 
\begin{equation}
C_{\Lambda },\text{ }\chi _{\Lambda },\text{ }S_{\Lambda },\text{ }%
A_{\Lambda }^{\mu },\text{ }\lambda _{\Lambda }^{\prime }=\lambda _{\Lambda
}+\frac{i}{2}\sigma ^{\mu }\partial _{\mu }\overline{{\large \chi }}%
_{\Lambda },\text{ }D_{\Lambda }^{\prime }=D_{\Lambda }+\frac{1}{2}\partial
^{2}C_{\Lambda }  \label{comp}
\end{equation}%
and varying the whole Lagrangian (\ref{ext}) with respect to these fields
one finds the constraints which appear to put on the $V$ superfield
components \cite{lec} 
\begin{equation}
C=C_{0},\text{ \ }\chi =0,\text{\ \ }A_{\mu }A^{\mu }=SS^{\ast }\text{.}
\label{const1}
\end{equation}%
They also determine the corresponding $D$-term (\ref{d}), $D=-P(C_{0}),$ for
the spontaneously broken supersymmetry. As usual, I only take a solution
with initial values for all fields (and their momenta) chosen so as to
restrict the phase space to vanishing values of the multiplier component
fields (\ref{comp}). This will provide a ghost-free theory with a positive
Hamiltonian.

Finally, implementing the constraints (\ref{const1}) into the total
Lagrangian $L_{tot}$ (\ref{ext}, \ref{lag3}) through the Lagrange multiplier
terms for component fields, we come to the emergent SUSY QED appearing in
the broken SUSY phase%
\begin{equation}
L^{\mathfrak{em}}=L_{SQED}+P(C)D\text{ }+\frac{D_{\Lambda }}{4}(C-C_{0})^{2}-%
\frac{C_{\Lambda }}{4}\left( A_{\mu }A^{\mu }-SS^{\ast }\right) \text{ .}
\label{fin}
\end{equation}%
The last two term with the component multiplier functions $C_{\Lambda }$ and 
$D_{\Lambda }$ of the auxiliary superfield $\Lambda $ (\ref{comp}) provide
the vacuum stability condition of the theory. In essence, one does not need
now to postulate from the outset gauge invariance for the physical SUSY QED
Lagrangian $L_{SQED}$. Rather, one can derive it following the GSE
conjecture (section 1.1) specified for Abelian theory. Indeed, due to the
constraints (\ref{const1}), the Lagrangian $L_{SQED}$ is only allowed to
have a conventional gauge invariant form 
\begin{equation}
L_{SQED}=-\text{ }\frac{1}{4}F^{\mu \nu }F_{\mu \nu }+i\lambda \sigma ^{\mu
}\partial _{\mu }\overline{\lambda }+\frac{1}{2}D^{2}  \label{444}
\end{equation}%
Thus, for the constrained vector superfield involved 
\begin{equation}
\widehat{V}(x,\theta ,\overline{\theta })=C_{0}+\frac{i}{2}\theta \theta S-%
\frac{i}{2}\overline{\theta }\overline{\theta }S^{\ast }-\theta \sigma ^{\mu
}\overline{\theta }A_{\mu }+i\theta \theta \overline{\theta }\overline{%
\lambda }-i\overline{\theta }\overline{\theta }\theta \lambda +\frac{1}{2}%
\theta \theta \overline{\theta }\overline{\theta }D,  \label{sup}
\end{equation}%
we have the almost standard SUSY QED Lagrangian with the same states - a
photon, a photino and an auxiliary scalar $D$ field - in its gauge
supermultiplet, while another auxiliary complex scalar field $S$ gets only
involved in the vector field constraint in (\ref{const1}). The linear
(Fayet-Iliopoulos) $D$-term with the effective coupling constant $P(C_{0})$
in (\ref{fin}) shows that supersymmetry in the theory is spontaneously
broken due to which the $D$ field acquires the VEV, $D=-P(C_{0})$. Taking
the nondynamical $S$ field in the constraint (\ref{const1}) to be some
constant background field we come to the SLIV\ constraint (\ref{const})
underlying the GSE conjecture. As is seen from this constraint in (\ref{fin}%
), one may only have the time-like SLIV in a SUSY framework but never the
space-like one. There also may be a light-like SLIV, if the $S$ field
vanishes\footnote{%
Indeed, this case, first mentioned in \cite{nambu}, may also mean
spontaneous Lorentz violation with a nonzero VEV $<A_{\mu }>$ $=(\widetilde{M%
},0,0,\widetilde{M})$ and Goldstone modes $A_{1,2}$ and $(A_{0}+A_{3})/2$\ $-%
\widetilde{M}.$ The "effective" Higgs mode $(A_{0}-A_{3})/2$ can be then
expressed through Goldstone modes so as the light-like condition $A_{\mu
}A^{\mu }=0$ to be satisfied.}. So, any possible choice for the $S$ field
corresponds to the particular gauge choice for the vector field $A_{\mu }$
in an otherwise gauge invariant theory. So, the massless photon appearing
first as a companion of a massless photino (being a Goldstone fermion in the
visible broken SUSY phase) remains massless due to this recovering gauge
invariance in the emergent SUSY QED. At the same time, the "built-in"
nonlinear gauge condition in (\ref{fin}) allows to treat the photon as a
vector Goldstone boson induced by an inactive SLIV.

\section{On \textbf{emergent \textbf{SUSY} Standard Models and GUTs}\ }

\subsection{Potential of Abelian and non-Abelian vector superfields}

Now, we extend our discussion to the non-Abelian internal symmetry case
given by some group $G$ with generators $t^{p}$ 
\begin{equation}
\lbrack t^{p},t^{q}]=if^{pqr}t^{r}\text{ , \ }Tr(t^{p}t^{q})=\delta ^{pq}%
\text{ }\ \text{(}p,q,r=0,1,...,\Upsilon -1\text{)\ }  \label{22}
\end{equation}%
where $f^{pqr}$ stand structure constants, while $\Upsilon $ is a dimension
of the $G$ group. This case may correspond in general to some Grand Unified
Theory which includes the Standard Model and its possible extensions. For
definiteness, I will be further focused on the $U(1)\times SU(N)$
symmetrical theories, though any other non-Abelian group in place of $SU(N)$
is also admissible. Such a split group form is dictated by the fact that in
the pure non-Abelian symmetry case supersymmetry does not get spontaneously
broken in a visible sector that makes it inappropriate for an outgrowth of
an emergence process\footnote{%
In principle, SUSY may be spontaneously broken in the visible sector even in
the pure non-Abelian symmetry case provided that the vector superfield
potential includes some essential high-dimension couplings.}. So, the theory
now contains the Abelian vector superfield $V$, as is given in (\ref{par1}),
and non-Abelian superfield multiplet $\boldsymbol{V}^{p}$ 
\begin{eqnarray}
\boldsymbol{V}^{p}(x,\theta ,\overline{\theta }) &=&\boldsymbol{C}%
^{p}+i\theta \boldsymbol{\chi }^{p}-i\overline{\theta }\overline{\boldsymbol{%
\chi }}^{p}+\frac{i}{2}\theta \theta \boldsymbol{S}^{p}-\frac{i}{2}\overline{%
\theta }\overline{\theta }\boldsymbol{S}^{\ast p}  \notag \\
&&-\theta \sigma ^{\mu }\overline{\theta }\boldsymbol{A}_{\mu }^{p}+i\theta
\theta \overline{\theta }\overline{\boldsymbol{\lambda }^{\prime }}^{p}-i%
\overline{\theta }\overline{\theta }\theta \boldsymbol{\lambda }^{\prime p}+%
\frac{1}{2}\theta \theta \overline{\theta }\overline{\theta }\boldsymbol{D}%
^{\prime p},  \label{par3}
\end{eqnarray}%
where its vector field components $\boldsymbol{A}_{\mu }^{p}$ are usually
associated with an adjoint gauge field multiplet, $(\boldsymbol{A}_{\mu
})_{j}^{i}\equiv (\boldsymbol{A}_{\mu }^{p}t^{p})_{j}^{i}$ \ ($%
i,j,k=1,2,...,N$ ; $p,q,r=1,2,...,N^{2}-1$). Note that, apart from the
conventional gaugino multiplet $\boldsymbol{\lambda }^{p}$ and the auxiliary
fields $\boldsymbol{D}^{p}$, the superfield $\boldsymbol{V}^{p}$ contains in
general the additional degrees of freedom in terms of the dynamical scalar
and fermion field multiplets $\boldsymbol{C}^{p}$ and $\mathbf{\chi }^{p}$
and nondynamical complex scalar field $\boldsymbol{S}^{p}$. Note that for
the non-Abelian superfield components I use hereafter the bold symbols and
take again the brief notations, $\boldsymbol{\lambda }^{\prime p}=%
\boldsymbol{\lambda }^{p}+\frac{i}{2}\sigma ^{\mu }\partial _{\mu }\overline{%
\boldsymbol{\chi }}^{p}$\ and $\boldsymbol{D}^{\prime p}=\boldsymbol{D}^{p}+%
\frac{1}{2}\partial ^{2}\boldsymbol{C}^{p}$.

Augmenting the SUSY and $U(1)\times SU(N)$ invariant GUT by some polynomial
potential of vector superfields $V$ and $\boldsymbol{V}^{p}$ one comes to 
\begin{equation}
\mathcal{L}=\mathcal{L}_{SGUT}+\frac{1}{2}D^{2}+\frac{1}{2}\boldsymbol{D}^{p}%
\boldsymbol{D}^{p}+[\xi V+b_{1}V^{3}/3+b_{2}V(\boldsymbol{VV})+b_{3}(%
\boldsymbol{VVV})/3]_{D}  \label{np2}
\end{equation}%
where $\xi $ and $b_{1,2,3}$ stand for coupling constants, and the last term
in (\ref{np2}) contains products of the Abelian superfield $V$ and the
adjoint $SU(N)$ superfield multiplet $\boldsymbol{V_{j}^{i}\equiv (%
\boldsymbol{V}^{p}}t\boldsymbol{^{p})_{j}^{i}}$. The round brackets denote
hereafter traces for the superfield $\boldsymbol{V_{j}^{i}}$ 
\begin{equation}
(\boldsymbol{VV...})\equiv Tr(\boldsymbol{VV...})\text{ }  \label{tra}
\end{equation}%
and its field components (see below). For simplicity, we restricted
ourselves to the third degree superfield terms in the Lagrangian $\mathcal{L}
$ to eventually have a theory at a renormalizible level. Furthermore, I have
only taken the odd power superfield terms that provides, as we see below, an
additional discrete symmetry of the potential with respect to\ the scalar
field components in the $V$ and $\boldsymbol{V}^{p}$ superfields%
\begin{equation}
C\rightarrow -\text{ }C,\text{ \ \ }\boldsymbol{C}^{p}\rightarrow -\text{ }%
\boldsymbol{C}^{p}.  \label{dis}
\end{equation}%
Finally, eliminating the auxiliary $D$ and $\boldsymbol{D}^{p}$ fields in
the Lagrangian $\mathcal{L}$ we come to the total potential for all
superfield bosonic field components written in terms of traces mentioned
above (\ref{tra}) 
\begin{eqnarray}
\mathcal{U}_{tot} &=&\mathcal{U}(C,\boldsymbol{C})+\frac{1}{2}\text{ }%
b_{1}C(A_{%
%TCIMACRO{\U{b5}}%
%BeginExpansion
{\mu}%
%EndExpansion
}A^{\mu }-S_{\alpha }S_{\alpha })+\frac{1}{2}\text{ }b_{2}C[(\boldsymbol{A}_{%
%TCIMACRO{\U{b5}}%
%BeginExpansion
{\mu}%
%EndExpansion
}\boldsymbol{A}^{\mu })-(\boldsymbol{S}_{\alpha }\boldsymbol{S}_{\alpha })] 
\notag \\
&&+\frac{1}{2}b_{2}[A_{%
%TCIMACRO{\U{b5}}%
%BeginExpansion
{\mu}%
%EndExpansion
}(\boldsymbol{A}^{\mu }\boldsymbol{C})-S_{\alpha }(\boldsymbol{S}_{\alpha }%
\boldsymbol{C})]+\frac{1}{2}b_{3}[(\boldsymbol{A}_{%
%TCIMACRO{\U{b5}}%
%BeginExpansion
{\mu}%
%EndExpansion
}\boldsymbol{A}^{\mu }\boldsymbol{C})-(\boldsymbol{S}_{\alpha }\boldsymbol{S}%
_{\alpha }\boldsymbol{C})]\text{ .}  \label{uuuu}
\end{eqnarray}%
Note that the potential terms depending only on scalar fields $C$ and $%
\boldsymbol{C_{j}^{i}\equiv (\boldsymbol{C}^{a}}t\boldsymbol{^{a})_{j}^{i}}$
are collected in 
\begin{equation}
\mathcal{U}(C,\boldsymbol{C})=\frac{1}{8}[\xi +b_{1}C^{2}+b_{2}(\boldsymbol{%
CC})]^{2}+\frac{1}{2}[b_{2}^{2}C^{2}(\boldsymbol{CC})+b_{2}b_{3}C(%
\boldsymbol{CCC})+\frac{1}{4}b_{3}^{2}(\boldsymbol{CCCC})]  \label{u'}
\end{equation}%
and complex scalar fields $S_{\alpha }$ and $\boldsymbol{S}_{\alpha }^{p}$ ($%
\alpha =1,2$) are now taken in the real field basis like as 
\begin{equation}
S_{1}=(S+S^{\ast })/2,\text{ \ \ }S_{2}=(S-S^{\ast })/2i\text{ ,}
\label{bas}
\end{equation}%
an so on. One can see that all these terms are invariant under the discrete
symmetry (\ref{dis}), whereas the vector field couplings in the total
potential $\mathcal{U}_{tot}$ (\ref{uuuu}) break it. However, they vanish
when the $V$ and $\boldsymbol{V}^{p}$ superfields are properly constrained
that we actually confirm in the next section.

Let us consider first the\ pure scalar field potential $\mathcal{U}$ (\ref%
{u'}). The \ corresponding extremum conditions for $C$ and\ $\boldsymbol{C}%
^{a}$ fields are, 
\begin{eqnarray}
\mathcal{U}_{C}^{\prime } &=&b_{1}(\xi +b_{1}C^{2})C+b_{2}(b_{1}-2b_{2})C(%
\boldsymbol{CC})=0,\text{ }  \notag \\
Tr(\mathcal{U}_{\boldsymbol{C_{j}^{i}}}^{\prime }) &=&3b_{2}C(\boldsymbol{CC}%
)+b_{3}(\boldsymbol{CCC})=0\text{ },  \label{extr}
\end{eqnarray}%
respectively. As shows the second partial derivative test, the simplest
solution to the above equations 
\begin{equation}
C_{0}=0\text{ , \ }\boldsymbol{C_{j}^{i}}=0  \label{sol3}
\end{equation}%
provides, under conditions put on the potential parameters,%
\begin{equation}
\xi ,\text{ }b_{1}>0\text{ ,\ }b_{2}\geq 0\text{ \ or \ }\xi ,\text{ }b_{1}<0%
\text{ ,\ }b_{2}\leq 0  \label{ccc}
\end{equation}%
its global minimum 
\begin{equation}
\mathcal{U}(C,\boldsymbol{C})_{\min }^{as}=\frac{1}{8}\xi ^{2}\text{ .}
\label{min2}
\end{equation}%
This minimum corresponds to the broken SUSY phase with the unbroken internal
symmetry $U(1)\times SU(N)$ that is just what one would want to trigger an
emergence process. This minimum appears in fact due to the Fayet-Iliopoulos
linear term in the superfield polynomial in (\ref{np2}). As can easily be
confirmed, in absence of this term, namely, for $\xi =0$ and any arbitrary
values of all other parameters, there is only the SUSY symmetrical solution
with unbroken internal symmetry%
\begin{equation}
\mathcal{U}(C,\boldsymbol{C})_{\min }^{sym}=0\text{ .}  \label{min3}
\end{equation}%
Interestingly, the symmetrical solution corresponding to the global minimum (%
\ref{min3}) may appear for the nonzero parameter $\xi $ as well 
\begin{equation}
\text{ }C_{0}^{(\pm )}=\pm \sqrt{-\xi /b_{1}}\text{, \ }\boldsymbol{C_{j}^{i}%
}=0  \label{sol4}
\end{equation}%
provided that 
\begin{equation}
\xi b_{1}<0\text{ .}  \label{cccc}
\end{equation}%
However, as we saw in the QED case, in the unbroken SUSY case one comes to
the trivial constant superfield when all factual constraints are included
into consideration \cite{lec} and, therefore, this case is in general of
little interest.

\subsection{Constrained vector supermultiplets}

Let us now take the vector fields $A_{\mu }$ and $\boldsymbol{A}_{\mu }^{p}$
into consideration that immediately reveals that, in contrast to the pure
scalar field part (\ref{u'}), $\mathcal{U}(C,\boldsymbol{C})$, the vector
field couplings in the total potential (\ref{uuuu}) make it unstable. This
happens, as was emphasized before, due the fact that bilinear term VEV
contributions of the vector fields $A_{\mu }$ and $\boldsymbol{A}_{\mu }^{p}$%
, as well as the auxiliary scalar fields $S_{\alpha }$ and $\boldsymbol{S}%
_{\alpha }^{p}$, are not properly compensated by appropriate four-linear
field terms which are generically absent in a supersymmetric theory
framework.

Again, as in the supersymmetric QED case considered above, the only possible
way to stabilize the ground state (\ref{sol3}, \ref{ccc}, \ref{min2}) seems
to seek the proper constraints on the superfields component fields ($C$, $%
\boldsymbol{C}^{p}$; $A_{\mu }$, $\boldsymbol{A}^{p}$; $S_{\alpha }$, $%
\boldsymbol{S}_{\alpha }^{p}$) themselves rather than on their expectation
values. Provided that such constraints are physically realizable, the
required vacuum will be automatically stabilized. This will be done again
through some invariant Lagrange multiplier couplings simply adding their $D$
terms to the above Lagrangian (\ref{np2}) 
\begin{equation}
\mathcal{L}_{tot}=\mathcal{L}+\frac{1}{2}\Lambda {\large (}V-C_{0})^{2}|_{D}+%
\frac{1}{2}\Pi (\boldsymbol{VV})|_{D}\text{ ,}  \label{ext1}
\end{equation}%
where $\Lambda (x,\theta ,\overline{\theta })$ and $\Pi (x,\theta ,\overline{%
\theta })$ are auxiliary vector superfields. Note that $C_{0}$ presented in
the first multiplier coupling is just the constant background value of the $%
C $ field for which the potential part $\mathcal{U}(C,\boldsymbol{C})$ in (%
\ref{uuuu}) vanishes as appears for the supersymmetric minimum (\ref{min3})
or has some nonzero value corresponding to the SUSY breaking minimum (\ref%
{min2}) in the visible sector.

I will consider both cases simultaneously using the same notation $C_{0}$
for either of the potential minimizing values of the $C$ field. The second
multiplier coupling in (\ref{ext1}) provides, as we will soon see, the
vanishing background value for the non-Abelian scalar field, $\boldsymbol{C}%
^{a}=0$, due to which the underlying internal symmetry $U(1)\times SU(N)$ is
left intact in both unbroken and broken SUSY phase. The Lagrange multiplier
terms presented in (\ref{ext1}) have in fact the simplest possible form that
leads to some nontrivial constrained superfields $V(x,\theta ,\overline{%
\theta })$ and $\boldsymbol{V}^{p}(x,\theta ,\overline{\theta })$. Writing
down their invariant $D$ terms through the component fields one finds the
precisely the same expression as in the SUSY QED \cite{lec} case for the
Abelian superfield $V$ and the slightly modified one for the non-Abelian
superfield $\boldsymbol{V}^{a}$ 
\begin{eqnarray}
\Pi (\boldsymbol{VV})|_{D} &=&C_{\Pi }\left[ \boldsymbol{CD}^{\prime }%
\boldsymbol{+}\left( \frac{1}{2}\boldsymbol{SS}^{\ast }\boldsymbol{-\chi
\lambda }^{\prime }\boldsymbol{-}\overline{\boldsymbol{\chi }}\overline{%
\boldsymbol{\lambda }^{\prime }}\boldsymbol{-}\frac{1}{2}\boldsymbol{A}_{\mu
}\boldsymbol{A}^{\mu }\right) \right]  \notag \\
&&+\text{ }\chi _{\Pi }\left[ 2\boldsymbol{C\lambda }^{\prime }\boldsymbol{+}%
i(\boldsymbol{\chi S}^{\ast }\boldsymbol{+}i\sigma ^{\mu }\overline{%
\boldsymbol{\chi }}\boldsymbol{A}_{\mu })\right] +\overline{\chi }_{\Pi }[2%
\boldsymbol{C}\overline{\boldsymbol{\lambda }^{\prime }}-i(\overline{%
\boldsymbol{\chi }}\boldsymbol{S}-i\boldsymbol{\chi \sigma }^{\mu }%
\boldsymbol{A}_{\mu })]  \notag \\
&&+\text{ }\frac{1}{2}S_{\Pi }\left( \boldsymbol{CS}^{\ast }\boldsymbol{+}%
\frac{i}{2}\overline{\boldsymbol{\chi }}\overline{\boldsymbol{\chi }}\right)
+\frac{1}{2}S_{\Pi }^{\ast }\left( \boldsymbol{CS-}\frac{i}{2}\boldsymbol{%
\chi \chi }\right)  \notag \\
&&+\text{ }2A_{\Pi }^{\mu }(\boldsymbol{CA}_{\mu }\boldsymbol{-\chi }\sigma
_{\mu }\overline{\boldsymbol{\chi }})+2{\large \lambda }_{\Pi }^{\prime }(%
\boldsymbol{C\chi })+2\overline{{\large \lambda }}_{\Pi }^{\prime }(%
\boldsymbol{C}\overline{\boldsymbol{\chi }})+\frac{1}{2}D_{\Pi }^{\prime }(%
\boldsymbol{CC})  \label{lm2}
\end{eqnarray}%
where the pairly grouped field bold symbols mean hereafter the $SU(N)$
scalar products of the component field multiplets (for instance, $%
\boldsymbol{CD}^{\prime }=\boldsymbol{C}^{p}\boldsymbol{D}^{\prime p}$, and
so forth) and 
\begin{equation}
C_{\Pi },\text{ }\chi _{\Pi },\text{ }S_{\Pi },\text{ }A_{\Pi }^{\mu },\text{
}\lambda _{\Pi }^{\prime }=\lambda _{\Pi }+\frac{i}{2}\sigma ^{\mu }\partial
_{\mu }\overline{{\large \chi }}_{\Pi },\text{ }D_{\Pi }^{\prime }=D_{\Pi }+%
\frac{1}{2}\partial ^{2}C_{\Pi }  \label{123}
\end{equation}%
are the component fields of the Lagrange multiplier superfield $\Pi
(x,\theta ,\overline{\theta })$ in the standard parametrization (\ref{par3}).

Varying the total Lagrangian (\ref{ext1}) with respect to the component
fields of both multipliers,\ (\ref{comp}) and (\ref{123}), and properly
combining their equations of motion we find the constraints which appear to
put on the $V$ and $\boldsymbol{V}^{a}$\ superfields components \cite{jlc} 
\begin{eqnarray}
C &=&C_{0}\text{,\ \ }\chi =0,\text{\ \ }A_{\mu }A^{\mu }=S_{\alpha
}S_{\alpha },\text{ \ }  \notag \\
\boldsymbol{C}^{p} &=&0,\text{ \ }\boldsymbol{\chi }^{p}=0,\text{ \ }(%
\boldsymbol{A}_{\mu }\boldsymbol{A}^{\mu })=(\boldsymbol{S}_{\alpha }%
\boldsymbol{S}_{\alpha })\text{ , \ }\alpha =1,2\text{ .}  \label{const4}
\end{eqnarray}%
As before in the SUSY QED case, one may only have the time-like SLIV in a
supersymmetric $U(1)\times SU(N)$ framework but never the space-like one
(there also may be a light-like SLIV, if the $S$ and $\boldsymbol{S}$ fields
vanish). Also note that we only take the solution with initial values for
all fields (and their momenta) chosen so as to restrict the phase space to
vanishing values of the multiplier component fields (\ref{comp}) and (\ref%
{123}) that will provide a ghost-free theory with a positive Hamiltonian.
Again, apart from the constraints (\ref{const4}), one has the equations of
motion for all fields involved in the basic superfields $V(x,\theta ,%
\overline{\theta })$ and $\boldsymbol{V}^{p}(x,\theta ,\overline{\theta })$.
With vanishing multiplier component fields (\ref{comp}) and (\ref{123}), as
was proposed above, these equations appear in fact as extra constraints on
components of the $V$ and $\boldsymbol{V}^{p}$ superfields. Indeed,
equations of motion for the $S_{\alpha }$, ${\large \chi }$ and $C$ fields,
on the one hand hand, and for the $\boldsymbol{S}_{\alpha }^{p},$ $%
\boldsymbol{\chi }^{p}$ and $\boldsymbol{C}^{p}$ fields, on the other, are
obtained by the corresponding variations of the total Lagrangian $\mathcal{L}%
^{tot}$ (\ref{ext1}) including the potential (\ref{uuuu}).

They are turned out to be, respectively, 
\begin{eqnarray}
S_{\alpha }C_{0} &=&0\text{ , \ }\lambda C_{0}=0\text{ ,\ \ }(\xi
+b_{1}C_{0}^{2})C_{0}=0\text{ , }  \notag \\
\boldsymbol{S}_{\alpha }^{p}C_{0} &=&0\text{ , }\boldsymbol{\lambda }%
^{p}C_{0}=0,\text{ }b_{2}[A_{%
%TCIMACRO{\U{b5}}%
%BeginExpansion
{\mu}%
%EndExpansion
}\boldsymbol{A}^{\mu }\boldsymbol{_{j}^{i}}-S_{\alpha }\boldsymbol{S}%
_{\alpha }\boldsymbol{_{j}^{i}}]+b_{3}[(\boldsymbol{A}_{%
%TCIMACRO{\U{b5}}%
%BeginExpansion
{\mu}%
%EndExpansion
}\boldsymbol{A}^{\mu })\boldsymbol{_{j}^{i}}-(\boldsymbol{S}_{\alpha }%
\boldsymbol{S}_{\alpha })\boldsymbol{_{j}^{i}}]=0  \label{nc6}
\end{eqnarray}%
where the basic constraints (\ref{const4}) emerging at the potential $%
\mathcal{U}(C,\boldsymbol{C})$ extremum point ($C_{0}$, $\boldsymbol{C}%
_{0}^{p}=0$) have been also used for both broken and unbroken SUSY case.
Note also that the equations for gauginos $\lambda $ and $\boldsymbol{%
\lambda }^{p}$ in (\ref{nc6}) are received by variation of the potential
terms in (\ref{np2}) containing fermion field couplings%
\begin{eqnarray}
\mathfrak{U} &=&b_{1}C(\chi \lambda ^{\prime }+\overline{\chi }\overline{%
\lambda ^{\prime }})+\text{ }b_{2}C[(\boldsymbol{\chi \lambda }^{\prime })+(%
\overline{\boldsymbol{\chi }}\overline{\boldsymbol{\lambda }^{\prime }})] 
\notag \\
&&+\frac{1}{2}b_{2}[\chi (\boldsymbol{\lambda }^{\prime }\boldsymbol{C})+%
\overline{\chi }(\overline{\boldsymbol{\lambda }^{\prime }}\boldsymbol{C}%
)+\lambda ^{\prime }(\boldsymbol{\chi C})+\overline{\lambda ^{\prime }}(%
\overline{\boldsymbol{\chi }}\boldsymbol{C})]  \notag \\
&&+b_{3}(\boldsymbol{\chi \lambda }^{\prime }\boldsymbol{C})+(\overline{%
\boldsymbol{\chi }}\overline{\boldsymbol{\lambda }^{\prime }}\boldsymbol{C})]%
\text{ .}  \label{fer}
\end{eqnarray}%
One can immediately see now that all equations in (\ref{nc6}) but the last
equation system turn to trivial identities in the broken SUSY case (\ref%
{sol3}) in which the corresponding $C$ field value appears to be identically
vanished, $C_{0}=0$. In the unbroken SUSY case (\ref{sol4}), this field
value \ is definitely nonzero, $C_{0}=\pm \sqrt{-\xi /b_{1}}$, and the
situation is radically changed. Indeed, as follows from the equations (\ref%
{nc6}), the auxiliary fields $S(x)$ and $\boldsymbol{S}^{p}$, as well as the
gaugino fields $\lambda (x)$ and $\boldsymbol{\lambda }^{p}(x)$ have to be
identically vanished. This causes in turn that the gauge vector fields field 
$A_{\mu }$ and $\boldsymbol{A}_{%
%TCIMACRO{\U{b5}}%
%BeginExpansion
{\mu}%
%EndExpansion
}^{p}$\ should also be vanished according to the basic constraints (\ref%
{const4}). So, we have to conclude, as in the SUSY QED case, that the
unbroken SUSY fails to provide stability of the potential (\ref{pot1a}) even
by constraining the superfields $V$ and $\boldsymbol{V}^{p}$ and, therefore,
only the spontaneously broken SUSY case could in principle lead to a
physically meaningful emergent theory.

\subsection{Broken SUSY phase: an emergent $U(1)\times SU(N)$ theory}

With the constraints (\ref{const4}) providing vacuum stability for the total
Lagrangian $\mathcal{L}_{tot}$ (\ref{ext1}) we eventually come to the
emergent theory with a local $U(1)\times SU(N)$ symmetry that appears in the
broken SUSY phase (\ref{sol3}). Actually, implementing these constraints
into the Lagrangian through the Lagrange multiplier terms for component
fields one has 
\begin{eqnarray}
\mathcal{L}^{\mathfrak{em}} &=&\mathcal{L}_{SGUT}+\frac{1}{2}\xi D\text{ }+%
\frac{D_{\Lambda }}{4}(C-C_{0})^{2}-\frac{C_{\Lambda }}{4}\left( A_{\mu
}A^{\mu }-SS^{\ast }\right)  \notag \\
&&+\frac{D_{\Pi }}{4}(\boldsymbol{CC})-\frac{C_{\Pi }}{4}\left( \boldsymbol{A%
}_{\mu }\boldsymbol{A}^{\mu }\boldsymbol{-SS}^{\ast }\right)  \label{laaag}
\end{eqnarray}%
with the multiplier component functions $C_{\Lambda }$ and $D_{\Lambda }$ of
the auxiliary superfield $\Lambda $ (\ref{comp}) and component functions $%
C_{\Pi }$ and $D_{\Pi }$ of the auxiliary superfield $\Pi $ (\ref{123})
presented in the Lagrangian (\ref{ext1}). Again, with these constraints and
the GSE conjecture (section 1.1) specified for non-Abelian theories, one
does not need to postulate gauge invariance for the physical SUSY GUT
Lagrangian $\mathcal{L}_{SGUT}$ from the outset. Instead, one can derive it
starting from an arbitrary relativistically invariant theory. Indeed, even
if the Lagrangian $\mathcal{L}_{SGUT}$ is initially taken to only possess
the global $U(1)\times SU(N)$ symmetry it will tend to uniquely acquire a
standard gauge invariant form 
\begin{eqnarray}
\mathcal{L}_{SGUT} &=&-\text{ }\frac{1}{4}F^{\mu \nu }F_{\mu \nu }+i\lambda
\sigma ^{\mu }\partial _{\mu }\overline{\lambda }+\frac{1}{2}D^{2}  \notag \\
&&-\text{ }\frac{1}{4}\boldsymbol{F}^{p\mu \nu }\boldsymbol{F}_{\mu \nu
}^{p}+i\boldsymbol{\lambda }^{p}\sigma ^{\mu }\mathcal{D}_{\mu }\overline{%
\boldsymbol{\lambda }}^{p}+\frac{1}{2}\boldsymbol{D}^{p}\boldsymbol{D}^{p}
\label{555}
\end{eqnarray}%
where the conventional gauge field strengths for both $U(1)$ and $SU(N)$
part and terms with proper covariant derivatives for gaugino fields $%
\boldsymbol{\lambda }^{p}$ necessarily appear \cite{jlc}. Again as in the
pure Abelian case, for the respectively constrained vector superfields $V$
and $\boldsymbol{V}^{p}$ we come in fact to a conventional SUSY GUT
Lagrangian with a standard gauge supermultiplet containing gauge bosons $%
A_{\mu }$ and $\boldsymbol{A}^{p}$, gauginos $\lambda $ and $\boldsymbol{%
\lambda }^{p}$, and auxiliary scalar $D$ and $\boldsymbol{D}^{p}$ fields,
whereas other auxiliary scalar fields $S_{\alpha }$ and $\boldsymbol{S}%
_{\alpha }^{p}$ get solely involved in the Lagrange multiplier terms (\ref%
{555}). Actually, the only remnant of the polynomial potential of vector
superfields $V$ and $\boldsymbol{V}^{p}$ (\ref{np2}) survived in the
emergent theory (\ref{laaag}) appears to be the Fayet-Iliopoulos $D$-term
which shows that supersymmetry in the theory is indeed spontaneously broken
and the $D$ field acquires the VEV, $D=-\frac{1}{2}\xi $.

Let us show now that this theory is in essence gauge invariant and the
constraints (\ref{const4})\ on the field space appearing due to the Lagrange
multiplier terms in (\ref{ext1}) are consistent with supersymmetry. Namely,
as was argued in \cite{lec} (see also \cite{jlc}), though constrained vector
superfield (\ref{sup}) in QED is not strictly compatible with the linear
superspace version of SUSY transformations, its supermultiplet structure can
be restored by appropriate supergauge transformations. Following the same
argumentation, one can see that similar transformations keep invariant the
constraints (\ref{const4}) put on the vector fields $A_{\mu }$ and $%
\boldsymbol{A}^{p}$. Leaving aside the $U(1)$ sector considered in \cite{lec}
in significant details, I will now focus on the $SU(N)$ symmetry case with
the constrained superfield $\boldsymbol{V}^{p}$ transformed as%
\begin{equation}
\boldsymbol{V}^{p}\rightarrow \boldsymbol{V}^{p}+\frac{i}{2}(\boldsymbol{%
\Omega -\Omega }^{\ast })^{p}  \label{ver}
\end{equation}%
The essential part of this transformation which directly acts on the vector
field constraint%
\begin{equation}
\boldsymbol{A}_{\mu }^{p}\boldsymbol{A}^{p\mu }=\boldsymbol{S}^{p}%
\boldsymbol{S}^{\ast p}  \label{vfc}
\end{equation}%
has the form 
\begin{equation}
\boldsymbol{V}^{p}\rightarrow \boldsymbol{V}^{p}+\frac{i}{2}\theta \theta 
\boldsymbol{F}^{p}-\frac{i}{2}\overline{\theta }\overline{\theta }%
\boldsymbol{F}^{\ast p}-\theta \sigma ^{\mu }\overline{\theta }\partial
_{\mu }\boldsymbol{\varphi }^{p}\text{ }  \label{trr}
\end{equation}%
where the real and complex scalar field components, $\boldsymbol{\varphi }%
^{p}$ and $\boldsymbol{F}^{p}$, in a chiral superfield parameter $%
\boldsymbol{\Omega }^{p}$ are properly activated. As a result, the
corresponding vector and scalar component fields, $\boldsymbol{A}_{\mu }^{p}$
and $\boldsymbol{S}_{\alpha }^{p}$, in the constrained supermultiplet $%
\boldsymbol{V}^{p}$ transform as 
\begin{equation}
\boldsymbol{A}_{\mu }^{p}\rightarrow \boldsymbol{a}_{\mu }^{p}=\boldsymbol{A}%
_{\mu }^{p}-\partial _{\mu }\boldsymbol{\varphi }^{p}\boldsymbol{,\ \ S}%
^{p}\rightarrow \boldsymbol{s}^{p}=\boldsymbol{S}^{p}+\boldsymbol{F}^{p}%
\text{ .}  \label{trrr}
\end{equation}

One can readily see that our basic Lagrangian $\mathcal{L}^{\mathfrak{em}}$ (%
\ref{laaag}) being gauge invariant and containing no the auxiliary scalar
fields $\boldsymbol{S}^{p}$ is automatically invariant under either of these
two transformations individually. In contrast, the supplementary vector
field constraint (\ref{vfc}), though it is also turned out to be invariant
under supergauge transformations (\ref{trrr}), but only if they act jointly.
Indeed, for any choice of the scalar $\boldsymbol{\varphi }^{p}$ in (\ref%
{trrr}) there can always be found such a scalar $\boldsymbol{F}^{a}$ (and
vice versa) that the constraint remains invariant. In other words, the
vector field constraint is invariant under supergauge transformations (\ref%
{trrr}) but not invariant under an ordinary gauge transformation. As a
result, in contrast to the Wess-Zumino case, the supergauge fixing in our
case will also lead to the ordinary gauge fixing. We will use this
supergauge freedom to reduce the scalar field bilinear $\boldsymbol{S}^{p}%
\boldsymbol{S}^{\ast p}$ to some constant background value and find a final
equation for the gauge function $\boldsymbol{\varphi }^{p}(x)$. It is
convenient to come to real field basis (\ref{bas}) for scalar fields $%
\boldsymbol{S}_{\alpha }^{p}$ and $\boldsymbol{F}_{\alpha }^{p}$ ($\alpha
=1,2$), and choose the parameter fields $\boldsymbol{F}_{\alpha }^{a}$ as 
\begin{equation}
\boldsymbol{F}_{\alpha }^{p}=r_{\alpha }\boldsymbol{\epsilon }^{p}(\mathbf{M}%
+\boldsymbol{f}),\boldsymbol{\ }r_{\alpha }\boldsymbol{s}_{\alpha }^{p}=0,%
\text{ \ }r_{\alpha }^{2}=1,\text{ }\boldsymbol{\epsilon }^{p}\boldsymbol{%
\epsilon }^{p}=1
\end{equation}%
so that the old $\boldsymbol{S}_{\alpha }^{p}$ fields in (\ref{trrr}) are
related to the new ones $\boldsymbol{s}_{\alpha }^{p}$ in the following way 
\begin{equation}
\boldsymbol{S}_{\alpha }^{p}=\boldsymbol{s}_{\alpha }^{p}-r_{\alpha }%
\boldsymbol{\epsilon }^{p}(\mathbf{M}+\boldsymbol{f}),\text{ }r_{\alpha }%
\boldsymbol{s}_{\alpha }^{p}=0,\text{ }\boldsymbol{S}_{\alpha }^{p}%
\boldsymbol{S}_{\alpha }^{p}=\boldsymbol{s}_{\alpha }^{p}\boldsymbol{s}%
_{\alpha }^{p}+(\mathbf{M}+\boldsymbol{f})^{2}\text{.}  \label{qq}
\end{equation}%
where $\mathbf{M}$ is a new mass parameter, $\boldsymbol{f}(x)$ is some
Higgs field like function, $r_{\alpha }$ is again the two-component unit
"vector" chosen to be orthogonal to the scalar $\boldsymbol{s}_{\alpha }^{p}$%
, while $\boldsymbol{\epsilon }^{p}$ is the unit $SU(N)$ adjoint vector.
This parametrization for the old fields $\boldsymbol{S}_{\alpha }^{p}$
formally looks as if they develop the VEV, $\left\langle \boldsymbol{S}%
_{\alpha }^{p}\right\rangle =$ $-r_{\alpha }\boldsymbol{\epsilon }^{p}%
\mathbf{M}$, due to which the related $SO(2)\times SU(N)$ symmetry would be
spontaneously violated and corresponding zero modes in terms of the new
fields $\boldsymbol{s}_{\alpha }^{p}$ could be consequently produced
(indeed, they they never appear in the theory). Eventually, for an
appropriate choice of \ the Higgs field like function $\boldsymbol{f}(x)$ in
(\ref{qq}) 
\begin{equation}
\boldsymbol{f}=-\mathbf{M}+\sqrt{\mathbf{M}^{2}-\boldsymbol{s}_{\alpha }^{p}%
\boldsymbol{s}_{\alpha }^{p}}  \label{w}
\end{equation}%
we come in (\ref{vfc}) to the condition 
\begin{equation}
\boldsymbol{A}_{\mu }^{p}\boldsymbol{A}^{p\mu }=\mathbf{M}^{2}\text{ .\ }
\label{111}
\end{equation}%
leading, as in the QED $U(1)$ symmetry case \cite{lec}, exclusively to the
time-like SLIV.

Remarkably, thanks to a generic high symmetry of the constraint (\ref{111})
one can apply the emergence conjecture with dynamically produced massless
gauge modes to any non-Abelian internal symmetry case as well, though SLIV
itself could produce only one zero vector mode. The point is that although
we only propose Lorentz invariance $SO(1,3)$ and internal symmetry $%
U(1)\times SU(N)$ of the Lagrangian $\mathcal{L}^{\mathfrak{em}}$ (\ref%
{laaag}), the emerged constraint (\ref{111}) possesses in fact a much higher
accidental symmetry $SO(\Upsilon ,3\Upsilon )$ determined by the dimension $%
\Upsilon =N^{2}-1$ of the $SU(N)$ adjoint representation to which the vector
fields $\boldsymbol{A}_{\mu }^{p}$ belong\footnote{%
Actually, a total symmetry even higher if one keeps in mind both constraints
(\ref{const}) and (\ref{111}) put on the vector fields $A_{\mu }$ and $%
\boldsymbol{A}_{\mu }^{a}$, respectively. As long as they are independent
the related total symmetry is in fact $SO(1,3)\times SO(\Upsilon ,3\Upsilon
) $ until it starts breaking.}. This symmetry is indeed spontaneously broken
at a scale $\mathbf{M}$ leading exclusively to the time-like SLIV case, as
is determined by the positive sign in the SUSY SLIV constraint (\ref{111}).
The emerging pseudo-Goldstone vector bosons may be in fact considered as
candidates for non-Abelian gauge fields which together with the true vector
Goldstone boson entirely complete the adjoint multiplet of the internal
symmetry group $SU(N)$. Remarkably, they remain strictly massless being
protected by the simultaneously generated non-Abelian gauge invariance. When
expressed in these zero modes, the theory look essentially nonlinear and
contains many Lorentz and CPT violating couplings. However, as in the SUSY
QED case, they do not lead to physical SLIV effects which due to
simultaneously generated gauge invariance appear to be strictly cancelled
out.

As in the pure QED case, one can calculate the gauge functions $\boldsymbol{%
\varphi }^{p}(x)$ comparing the relation between the old and new vector
fields in (\ref{trrr}) with a conventional SLIV parametrization for
non-Abelian vector fields \cite{jlc} 
\begin{equation}
\text{\ \ }\boldsymbol{A}_{\mu }^{p}=\boldsymbol{a}_{\mu }^{p}+\boldsymbol{n}%
_{\mu }^{p}\sqrt{\mathbf{M}^{2}-\boldsymbol{n}^{2}\boldsymbol{a}^{2}}\text{ }%
,\text{ \ }\boldsymbol{n}_{\mu }^{p}\boldsymbol{a}^{p\mu }\text{\ }=0\text{
\ \ \ }(\boldsymbol{a}^{2}\equiv \boldsymbol{a}_{\mu }^{p}\boldsymbol{a}%
^{p\mu }).  \label{supp}
\end{equation}%
They are expressed through the non-Abelian Goldstone and pseudo-Goldstone
modes $\boldsymbol{a}_{\mu }^{p}$ 
\begin{equation}
\boldsymbol{\varphi }^{p}=\boldsymbol{\epsilon }^{p}\int^{x}d(n_{\mu }x^{\mu
})\sqrt{\mathbf{M}^{2}-\boldsymbol{n}^{2}\boldsymbol{a}^{2}}\text{ .}
\end{equation}%
Here $n_{\mu }$ is the unit Lorentz vector being analogous to the vector
introduced in the Abelian case (\ref{vev1}), which is now oriented in
Minkowskian spacetime so as to be "parallel" to the vacuum unit $\boldsymbol{%
n}_{\mu }^{p}$ matrix. This matrix can be taken in the "two-vector" form 
\begin{equation}
\boldsymbol{n}_{\mu }^{p}=n_{\mu }\boldsymbol{\epsilon }^{p}\text{ },\text{ }%
\boldsymbol{\epsilon }^{p}\boldsymbol{\epsilon }^{p}=1  \label{vec}
\end{equation}%
where $\boldsymbol{\epsilon }^{p}$ is the unit $SU(N)$ group vector
belonging to its adjoint representation.

\subsection{Some immediate outcomes}

Quite remarkably, an obligatory split symmetry form $U(1)\times SU(N)$ (or $%
U(1)\times G$, in general) of plausible emergent theories which could exist
beyond the prototype QED case, leads us to the standard electroweak theory
with an $U(1)\times SU(2)$ symmetry as the simplest possibility. The
potential of type (\ref{np2}) written for the corresponding superfields
requires spontaneous SUSY breaking in the visible sector to avoid the vacuum
instability in the theory. Eventually, this requires the SLIV type
constraints to be put on the hypercharge and weak isospin vector fields,
respectively,%
\begin{equation}
B_{\mu }B^{\mu }=\mathrm{M}^{2}\text{ , \ }\boldsymbol{W}_{\mu }^{p}%
\boldsymbol{W}^{p\mu }=\mathbf{M}^{2}\text{ \ }(p=1,2,3).  \label{ew}
\end{equation}%
These constraints are independent from each other and possess, as was
generally argued above, the total symmetry $SO(1,3)\times SO(3,9)$ which is
much higher than the actual Lorentz invariance and electroweak $U(1)\times
SU(2)$ symmetry in the theory. Thanks to this fact, one Goldstone and three
pseudo-Goldstone zero vector modes $b_{\mu }$ and $\boldsymbol{w}_{\mu }^{p}$
are generated to eventually complete the gauge multiplet of the Standard
Model%
\begin{eqnarray}
B_{\mu } &=&b_{\mu }+n_{\mu }\sqrt{\mathrm{M}^{2}-b_{\mu }b^{\mu }}\text{ ,
\ }n_{\mu }b_{\mu }=0\text{ ,}  \notag \\
\boldsymbol{W}_{\mu }^{p} &=&\boldsymbol{w}_{\mu }^{p}+n_{\mu }\boldsymbol{%
\epsilon }^{p}\sqrt{\mathbf{M}^{2}-\boldsymbol{w}_{\mu }^{q}\boldsymbol{w}%
^{q\mu }}\text{ , }n_{\mu }\boldsymbol{w}^{p\mu }=0\text{ }  \label{ew1}
\end{eqnarray}%
where the unit vectors $n_{\mu }$ and $\boldsymbol{\epsilon }^{p}$ are
defined in accordance with a rectangular unit matrix $\boldsymbol{n}_{\mu
}^{p}$ taken in the two-vector form (\ref{vec}). The true vector Goldstone
boson appear to be some superposition of the zero modes $b_{\mu }$ and $%
\boldsymbol{w}_{\mu }^{3}$. This superposition is in fact determined by the
conventional Higgs doublet in the model since just through the Higgs field
couplings these modes are only mixed \cite{par}. Thus, when the electroweak
symmetry gets spontaneously broken an accidental degeneracy related to the
total symmetry of constraints mentioned above is lifted. As a consequence,
the vector pseudo-Goldstones acquire masses and only photon, being the true
vector Goldstone boson in the model, is left massless. In this sense, there
is not much difference for a photon in emergent QED and SM: it emerges as a
true vector Goldstone boson in both frameworks.

Going beyond the Standard Model we unavoidably come to the flipped $SU(5)$
GUT \cite{fl} as a minimal and in fact distinguished possibility. Indeed,
the $U(1)$ symmetry part being mandatory for emergent theories now naturally
appears as a linear combination of a conventional electroweak hypercharge
and another hypercharge belonging to the standard $SU(5).$ The flipped $%
SU(5) $ GUT has several advantages over the standard $SU(5)$ one: the
doublet-triplet splitting problem is resolved with use of only minimal Higgs
representations and protons are naturally long lived, neutrinos are
necessarily massive, and supersymmetric hybrid inflation can easily be
implemented successfully. Also in string theory, the flipped $SU(5)$ model
is of significant interest for a variety of reasons. In essence, the
above-mentioned natural solution to the doublet-triplet splitting problem
without using large GUT representations is in the remarkable conformity with
string theories where such representations are typically unavailable. Also,
in weakly coupled heterotic models, the flipped $SU(5)$ allows to achieve
gauge coupling unification at the string scale $10^{17}$ \textrm{GeV }if
some extra vector-like particles are added. They are normally taken to
transform in the $10$ and $\overline{10}$ representations, that is easy to
engineer in string theory.

So, supersymmetric emergent theories look attractive both theoretically and
phenomenologically whether they are considered at low energies in terms of
the Standard Model or at high energies as the flipped $SU(5)$ GUTs being
inspired by superstrings.

\section{Summary}

As we have seen above, spontaneous Lorentz violation in a vector field
theory framework may be active as in the composite and potential-based
models leading to physical Lorentz violation, or inactive as in the
constraint-based models resulting in the nonlinear gauge choice in an
otherwise Lorentz invariant theory. Remarkably, between these two basic SLIV
versions SUSY unambiguously chooses the inactive SLIV case. Indeed, SUSY
theories only admit the bilinear mass term in the vector field potential
energy. As a result, without a stabilizing quartic vector field terms, the
physical spontaneous Lorentz violation never occurs in SUSY theories. Hence
it follows that the composite and potential-based SLIV models can in no way
be realized in the SUSY context. This may have far-going consequences in
that supergravity and superstring theories could also disfavor such models
in general.

Though, even in the case when SLIV is not physical it inevitably leads to
the generation of massless photons as vector NG bosons provided that SUSY
itself is spontaneously broken. In this sense, a generic trigger for
massless photons to dynamically emerge happens to be spontaneously broken
supersymmetry rather than physically manifested Lorentz noninvariance. To
see how this idea might work we have considered supersymmetric QED model
extended by an arbitrary polynomial potential of a general vector superfield
that induces spontaneous SUSY violation in the visible sector, and gauge
invariance gets broken as well. \ Nevertheless, the special gauge invariance
is in fact recovered in the broken SUSY phase that universally protects the
photon masslessness.

All basic arguments developed in SUSY QED were then generalized to Standard
Model and Grand Unified Theories. Remarkably, thanks to a generic high
symmetry of the length-fixing SLIV constraint (\ref{111}) put on the vector
fields the emergence conjecture with dynamically produced massless gauge
modes can be applied to any non-Abelian global internal symmetry case due to
which it gets converted into to the local one. For definiteness, we have
focused above on the $U(1)\times SU(N)$ symmetrical theories. Such a split
group form is dictated by the fact that in the pure non-Abelian symmetry
case one only has the SUSY invariant phase in the theory that would make it
inappropriate for an outgrowth of an emergence process. As we briefly
discussed, supersymmetric emergent theories look attractive both
theoretically and phenomenologically whether they are considered at low
energies in terms of the Standard Model or at high energies as the flipped $%
SU(5)$ GUTs inspired by superstrings.

However, their most generic manifestations, as I discussed here in Bled
about a year ago \cite{lec} (for more details, see also \cite{jlc}), is
related to a spontaneous SUSY violation in the visible sector that seems to
open a new avenue for exploring the origin of gauge symmetries. Indeed, the
photino emerging due to this violation will be then mixed with another
goldstino which stems from a spontaneous SUSY violation in the hidden
sector. Eventually, it largely turns into light pseudo-goldstino whose
physics seems to be of special interest. Such pseudo-Goldstone photinos
might appear typically as the $\mathrm{eV}$ scale stable LSP or the
electroweak scale long-lived NLSP, being accompanied by a very light
gravitinos in both cases. Their observation could shed some light on an
emergence nature of gauge symmetries.

\section*{ Acknowledgments}

Discussions with the participants of the Workshop \textquotedblright What
Comes Beyond the Standard Models?\textquotedblright\ (21--28 July 2014,
Bled, Slovenia) were very useful. This work is partially supported by
Georgian National Science Foundation (Contracts No. 31/89 and No.
DI/12/6-200/13).

\end{document}